\documentclass[prl,twocolumn,superscriptaddress]{revtex4-2}
\usepackage{graphicx}
\usepackage{latexsym}
\usepackage{amsmath}
\usepackage{amsfonts}
\usepackage{amssymb}
\usepackage{bm}
\usepackage{txfonts}
\newcommand{\fig}[2]{\includegraphics[width=#1]{#2}}
\newcommand{{\sr}}{{Sr$_2$IrO$_4$}}
\newcommand{\bp}{{\textbf{p}}}
\newcommand{\bZ}{{\textbf{Z}}}
\newcommand{\tUVG}{{$t$-$U$-$V$-$\Gamma_2$}}

\newcommand{\ttai}{${\bm \tau}$}

\begin{document}
\title{Microscopic model realization of $\bm d$-wave pseudospin current order in Sr$_{\bm 2}$IrO$_{\bm 4}$ }

\author{Jin-Wei Dong}
\affiliation{CAS Key Laboratory of Theoretical Physics, Institute of Theoretical Physics, Chinese Academy of Sciences, Beijing 100190, China}
\affiliation{School of Physical Sciences, University of Chinese Academy of Sciences, Beijing 100049, China}
\affiliation{Anhui Province Key Laboratory of Condensed Matter Physics at Extreme Conditions, High Magnetic Field Laboratory, Chinese Academy of Sciences, Hefei 230031, China}

\author{Yun-Peng Huang}
\affiliation{Beijing National Laboratory for Condensed Matter Physics and Institute of Physics, Chinese Academy of Sciences, Beijing 100190, China}

\author{Ziqiang Wang}
\thanks{wangzi@bc.edu}
\affiliation{Department of Physics, Boston College, Chestnut Hill, MA 02467, USA}

\author{Sen Zhou}
\thanks{zhousen@itp.ac.cn}
\affiliation{CAS Key Laboratory of Theoretical Physics, Institute of Theoretical Physics, Chinese Academy of Sciences, Beijing 100190, China}
\affiliation{School of Physical Sciences, University of Chinese Academy of Sciences, Beijing 100049, China}
\affiliation{CAS Center for Excellence in Topological Quantum Computation, University of Chinese Academy of Sciences, Beijing 100049, China}

\begin{abstract}
The $d$-wave pseudospin current order ($d$PSCO) with staggered circulating pseudospin current has been proposed as the hidden electronic order to describe the unexpected breaking of spatial symmetries in stoichiometric \sr\ and the unconventional pseudogap phenomena in electron doped \sr.
However, a microscopic model for the emergence of $d$PSCO is still lacking.
The nearest neighbor Coulomb repulsion $V$, which is expected to be significant in \sr\ due to the large spatial extension of the Ir $5d$ orbitals, is capable of driving $d$PSCO on the mean-field level, albeit the latter is energetically degenerate to the staggered flux phase with circulating charge current.
We find the in-plane anisotropy $\Gamma_2$ in the effective superexchange interaction between $J_\text{eff}={1\over 2}$ pseudospins, originating from the cooperative interplay between Hund's rule coupling and spin-orbit coupling of Ir $5d$ electrons, is able to lift the degeneracy and stabilize the pseudospin currents.
The effective single-orbital model of $J_\text{eff}={1\over 2}$ electrons, including onsite Coulomb repulsion $U$, nearest neighbor Coulomb repulsion $V$, and the in-plane anisotropy $\Gamma_2$, is then studied.
We obtain the mean-field ground states, analyze their properties, and determine the phase diagram of stoichiometric \sr\ in the plane spanned by $U$ and $V$ at a fixed $\Gamma_2$.
We demonstrate the realization of $d$PSCO, as well as its competition and coexistence with antiferromagnetism.
Remarkably, we find the coexistence of $d$PSCO and antiferromagnetism naturally leads to spin bond nematicity, with the spin directions of these three orders forming nontrivial chirality.
Furthermore, we show that the emergence of the coexistent state and its chirality can be tuned by carrier doping.
\end{abstract}
\maketitle

\section{I. Introduction}

The layered square-lattice iridate \sr\ has recently attracted much attention partly due to its close resemblance to the high-temperature cuprate superconductors~\cite{Kim2008prl,Kim2009sci, Pesin2010natphy, Krempa2014annurev, Rau2016annurev, Schaffer2016reprogre, DaiJiXia2014prb, Winter2017conmat, Bertinshaw2019annurev, Lenz2019condmatt, LuChengliang2020advmat, zwartsenberg2020natphy}.
It is isostructural to La$_2$CuO$_4$ and the stoichiometric \sr\ becomes a canted antiferromagnetic (AFM) insulator below the N\'eel temperature $T_N\simeq$ 230 K~\cite{Kim2008prl, Kim2009sci}.
The magnetic excitations are well described by pseudospin-$1\over 2$ Heisenberg model on the square lattice, with strong AFM exchange coupling $J\simeq 60$ meV~\cite{Kim2012prl, Calder2018prb}.
This is believed to be the essential physics of the cuprates and thus naturally leads to the expectation that \sr\ can be another platform for unconventional high-temperature superconductivity upon carrier doping~\cite{WangFa2011prl, Watanabe2013prl, WangQiangHua2014prb, MengZiYang2014prl}.
Although there is not yet firm evidence for superconductivity, a remarkable range of cuprate phenomenology has been observed in electron- and hole-doped \sr, including Fermi surface pockets~\cite{Torre2015prl}, Fermi arcs~\cite{Kim2014sic}, pseudogaps~\cite{FengDL2015prx, Battisti2017natphy, Peng2022npj}, and V-shaped tunneling spectra that potentially signals $d$-wave superconductivity~\cite{Kim2016natphy, ZhaoHe2019natphy}.

At stoichiometry, neutron and resonant X-ray measurements reveal that the magnetic moments in the canted AFM insulator are aligned in the basal $ab$ plane, with their directions tracking the staggered IrO$_6$ octahedra rotation about the $c$ axis due to strong spin-orbit coupling~\cite{Boseggia2013condmatt, Crawford1994prb, YeFeng2013prb, Dhital2013prb, Torchinsky2015prl}.
The resulting net ferromagnetic moment of each layer is shown to order in a $+--+$ pattern along the $c$ axis~\cite{Kim2009sci,Boseggia2013prl}.
This magnetic ground state belongs to a centrosymmetric orthorhombic magnetic point group $2/m1'$ with spatial $C_{2z}$ rotation, inversion, and time-reversal symmetries (TRS) \cite{Zhao2016natphy,Norman2016prb}.
Recent optical second-harmonic generation experiments~\cite{Zhao2016natphy,Seyler2020prb}, however, reported evidence of unexpected breaking of spatial rotation and inversion symmetries, pointing to the existence of a symmetry-breaking hidden order.
It is further supported by polarized neutron diffraction~\cite{Jeong2017natcomm} and muon spin relaxation measurements~\cite{Tan2020prb} which revealed the breaking of TRS.
Intriguingly, magnetic resonant X-ray scattering measurements conducted on the electron-doped \sr\ have uncovered a unidirectional spin density wave in the pseudogap phase~\cite{ChenXiang2018natcomm}, further supporting the idea that the pseudogap is associated with a symmetry-breaking hidden order.
It was argued that the broken symmetries can be caused by loop currents~\cite{Zhao2016natphy,Jeong2017natcomm,Tan2020prb,Murayama2021prx} which were proposed to account for the pseudogap physics in cuprates~\cite{Varma1997prb, Varma2006prb, Varma2014conden}.
However, the oxygen 2$p$ states in \sr\ are much further away from the Fermi level than those in the cuprates~\cite{Kim2008prl, Moon2006prb}, making it disadvantageous to develop the loop currents that requires low-energy oxygen 2$p$ states.

The $d$-wave pseudospin current order ($d$PSCO), with pseudospin-up electrons staggered circulating along one direction and pseudospin-down electrons in the opposite direction, has been proposed as an alternative candidate for the hidden electronic order in \sr~\cite{ZhouSen2017prx}.
Symmetry analysis shows that the coexistence of $d$PSCO and canted AFM with a particular $c$-axis stacking pattern has the symmetries consistent with all available experimental observations on the stoichiometric \sr\ below the N\'eel temperature~\cite{ZhouSen2021prb}.
This coexistent phase is a magnetoelectric state that breaks twofold rotation, spatial inversion, and TRS~\cite{ZhouSen2021prb}.
In addition, it can account for the observed splitting of bands~\cite{Torre2015prl} at $(\pi, 0)$ whose twofold degeneracy is otherwise protected by certain lattice symmetries~\cite{ZhouSen2017prx, HanJW2020prb, Kim2021prb}.
Upon sufficient electron doping such that the magnetism is completely suppressed, $d$PSCO produces Fermi pockets and Fermi arcs in the nonmagnetic electron-doped \sr\ \cite{ZhouSen2017prx}, in good agreement with the pseudogap phenomena revealed by angle-resolved photoemission and scanning tunneling microscopy measurements~\cite{Torre2015prl,Kim2014sic,FengDL2015prx,Kim2016natphy}.
While describing remarkably well the unexpected symmetry properties and the unconventional quasiparticle behaviours observed in both stoichiometric and electron-doped \sr, the physical origin of $d$PSCO is unclear and a microscopic model for its emergence is still lacking.

\begin{figure}
\begin{center}
\fig{3.4in}{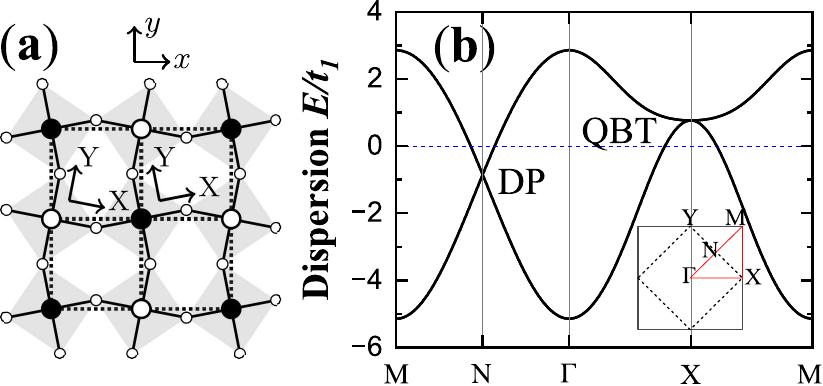}
\caption{(a) Schematic picture of one IrO$_2$ layer. Large filled or open circles denote the Ir atoms on the two sublattices, and small open circles are oxygens. Lowercase $x$, $y$  and capital $X$, $Y$ indicate, respectively, the global and sublattice-dependent local cubic axis.
(b) Tight-binding band structure of the free electrons, displaying a DP at $N$ and a QBT at $X$ point. The inset in (b) shows the one-Ir BZ (solid black lines), the reduced BZ (dotted black lines), and the high-symmetry points labeled by $\Gamma=(0, 0)$, $X=(\pi, 0)$, $Y=(0, \pi)$, $M=(\pi, \pi)$, and $N=(\pi/2, \pi/2)$.} \label{fig1}
\end{center}
\end{figure}

In this work, we discuss the emergence of $d$PSCO in an effective single-orbital model for pseudospin-$1\over 2$ electrons of \sr\ in the local basis (see Fig. \ref{fig1}a) tracking the staggered IrO$_6$ octahedra rotation, in which the canted AFM becomes a perfect Ne\'el order~\cite{WangFa2011prl,Jackeli2009prl}.
Hereinafter, we replace $d$PSCO by $d$SCO ($d$-wave spin current order) for convenience.
Due to the large spatial extension of the Ir $5d$ orbitals, the off-site Coulomb repulsions are significant in \sr\ and expected to play important role in the development of $d$SCO.
Indeed, it has been shown on half-filled honeycomb lattice that off-site Coulomb repulsions can produce $d$SCO with staggered spin current on the mean-field level~\cite{Raghu2008prl}.
It has the continuous global SO(3) symmetry associated with the rotation of the spin direction, and energetically degenerate with the staggered flux phase (SFP) with circulating charge current, leading to spin Hall effect and anomalous Hall effect, respectively.
TRS is broken in SFP but preserved in $d$SCO, and thus these two states can never coexist.
It is argued that quantum fluctuations lifts the degeneracy and favors spin current over charge current~\cite{Raghu2008prl}.

Interestingly, we note that the in-plane anisotropy $\Gamma_2$ in the effective superexchange interactions between the $J_\text{eff} ={1\over 2}$ pseudospins, originating from the cooperative interplay between Hund's rule coupling and spin-orbit coupling of the Ir 5$d$ electrons~\cite{Jackeli2009prl}, can lift the degeneracy between SFP and $d$SCO within mean-field theories.
Furthermore, it breaks the SO(3) rotation symmetry of $d$SCO down to $C_{4z}$ by orientating the spins of $d$SCO along one of the four easy axes, i.e., $[\pm 1,\pm1,0]$.
This motivates us to investigate the realization of $d$SCO in a concrete effective single-orbital \tUVG\ model of pseudospin-$1\over 2$ electrons for \sr, where $t$ denotes the kinetic hoppings, $U$ for on-site Coulomb repulsion, $V$ for nearest-neighbor (nn) Coulomb repulsion, and $\Gamma_2$ for the in-plane anisotropy of pseudospins.

We discuss the emergence of $d$SCO in the \tUVG\ model, and investigate its competition and coexistence with AFM.
The rest of the paper is organized as follows.
Sec. II introduces the effective single-orbital \tUVG\ model for pseudospin-$1\over 2$ electrons in \sr.
The onsite Coulomb repulsion $U$ is treated by SU(2) spin-rotation invariant slave-boson mean-field theory, while interactions on nn bonds, $V$ and $\Gamma_2$, are mean-field decoupled into bond channels.
In Sec. III, the \tUVG\ model is solved self-consistently at half-filling for stoichiometric \sr.
We obtain the mean field ground states, analyze their properties, and determine the phase diagram at a fixed nonzero $\Gamma_2$.
The phase diagram consists of paramagnetic (PM), AFM, $d$SCO, and the coexistent state involving the latter two.
We note that the coexistent state breaks the $C_{4z}$ rotation symmetry of $J_\text{eff} ={1\over 2}$ pseudospins, and naturally leads to spin bond nematicity (sBN) with nontrivial chirality.
We then study the doping evolution of these states and illustrate that the emergence of coexistent state and its chirality can be tuned by the carrier doping.
Possible connections to experimental observations are also discussed.
Summaries are presented in Sec. IV.

\section{II. Model and Method}
\textit{The \tUVG\ model}. We start with an effective square lattice single-orbital \tUVG\ model for the pseudospin-$1\over 2$ electrons in the local basis depicted in Fig. \ref{fig1}a that tracks the staggered IrO$_6$ octahedra rotation in \sr,
\begin{align}
\hat{H} =& - \sum_{i j, \alpha} t_{i j} c_{i \alpha}^{ \dag} c_{j \alpha}+ U \sum_i \hat{n}_{i \uparrow} \hat{n}_{i \downarrow} \label{Ham} \\
& +V \sum_{\langle i j \rangle} \hat{n}_i \hat{n}_j + \Gamma_2 \sum_{\langle i j \rangle} \tau_{ij} (\hat{S}^x_i \hat{S}^x_j - \hat{S}^y_i \hat{S}^y_j), \nonumber
\end{align}
where $\tau_{ij}=(- 1)^{i_y + j_y}$ is the standard $d$-wave form factor on nn bonds, $c_{i \alpha}^{\dag}$ creates a pseudospin-$\alpha$ ($\alpha=\uparrow, \downarrow$) electron at site $i$, the density operators $\hat{n}_{i\alpha} = c^\dagger_{i\alpha} c_{i\alpha}$, $\hat{n}_i=\sum_\alpha \hat{n}_{i\alpha}$, and the pseudospin operator $\hat{S}_i^{\eta} = \frac{1}{2} \sum_{\alpha \beta} c^{\dag}_{i \alpha} \sigma^{\eta}_{\alpha \beta} c_{i \beta}$ with $\sigma^\eta$ the Pauli matrices for $\eta = x, y, z$.
To describe the low-energy quasiparticle dispersion (Appendix A), the hopping parameters are chosen according to $t_{ij}$ = $(t_1, t_2, t_3)$ = $(218, 52, - 18)$ meV for the first, second, and third nn hopping, respectively.
Fig. \ref{fig1}b shows the tight-binding dispersion along the high-symmetry path in the reduced Brillouin zone (BZ), displaying the characteristic Dirac point (DP) at $N$ and quadratic band touching (QBT) at $X$ point.
For this set of band parameters, the QBT is higher in energy than DP by $4t_2-8t_3=352$ meV.
In Eq. (\ref{Ham}), $U$ and $V$ are the onsite and nn Coulomb repulsions, and $\Gamma_2$ is the in-plane anisotropy for $J_\text{eff}={1\over 2}$ pseudospins.
It has been shown that $\Gamma_2$ is generated by the cooperative interplay between Hund's rule coupling and spin-orbit coupling of Ir 5$d$ electrons, and induces a small in-plane magnetic gap via quantum fluctuations~\cite{Jackeli2009prl}.

\textit{Mean-field theories}. To treat the onsite interaction nonperturbatively and consider in-plane magnetic order conveniently, we use the Kotliar-Ruckenstein slave-boson formulation with SU(2) spin-rotation invariance~\cite{Wolfle1989prb,Fresard1992modphy,JiangKun2014prb,Huang2020prb}, in which the physical electron operator is written as
\begin{equation}
c_\alpha = Z_{\alpha \beta} f_\beta, \quad
\bZ = \textbf{L}^{-1/2}  \left( e^\dagger \bp + \bar{\bp}^\dagger d \right) \textbf{R}^{-1/2},
\end{equation}
where $f_\alpha$ is a spin-$1\over 2$ fermion operator, and $e$, $d$, and $\bp$ are boson operators describe the holon, doublon, and singly occupied sites.
The SU(2) spin-rotation invariance is achieved by the $2\times 2$ matrix representation of the singly occupied site, \textbf{p}, with element $p_{\alpha\beta} = {1\over \sqrt{2}} \sum_{\mu=0,x,y,z} p_\mu \sigma^\mu_{\alpha\beta}$, and its time-reversal transformation $\bar{\bp} = \hat{T} \bp \hat{T}^{-1}$.
The $2\times 2$ matrix operator $\textbf{L} =(1-d^\dagger d) \sigma^0 -\bp^\dagger \bp$, $\textbf{R} = (1-e^\dagger e) \sigma^0 -\bar{\bp}^\dagger \bar{\bp}$, with $\sigma^0$ the $2\times 2$ identity matrix.
This form ensures the spin rotation invariance and the correct noninteracting limit within the mean-field approximation~\cite{JiangKun2014prb}.
The intersite interactions on nn bonds, $V$ and $\Gamma_2$, are rewritten in terms of Hubbard-Stratonivich fields $\hat{\chiup}^\mu_{\langle ij \rangle} =\sum_{\alpha\beta} f^\dagger_{i\alpha} \sigma^\mu_{\alpha\beta} f_{j\beta}$, $\mu=0,x,y,z$~\cite{Raghu2008prl}.
The charge density fields corresponding to the direct Hartree decoupling of $V$ are neglected to avoid double-counting, since their contribution is already included in the LDA \cite{Jiang2016prb}.
Furthermore, it is easy to show that the spin density fields for the in-plane anisotropy $\Gamma_2$ are irrelevant in the two-sublattice states considered in this work.
As a result, the Hamiltonian becomes
\begin{align}
\widetilde{H}=&- \sum_{ij, \alpha\beta\gamma} t_{ij}
f^\dagger_{i \alpha} Z^\dagger_{i,\alpha \gamma} Z_{j,\gamma\beta} f_{j \beta} +U \sum_i d^\dagger_i d_i +\sum_i \lambda_i \hat{Q}_i  \label{Hsb} \\
+ & \sum_{i,\mu} \lambda^\mu_i \hat{Q}^\mu_i -\sum_{\langle ij \rangle} \left\{ \frac{V}{2} \sum_{\mu} \big| \hat{\chiup}^\mu_{ij} \big|^2+ \frac{\Gamma_2}{4} \tau_{ij} \left[ \big| \hat{\chiup}^x_{ij} \big|^2 - \big|\hat{\chiup}^y_{ij} \big|^2 \right] \right\}, \nonumber
\end{align}
where $\lambda_i$ and $\lambda^\mu_i$ $(\mu=0,x,y,z)$ are Lagrange multipliers introduced to enforce the local constraints for the completeness of the Hilbert space
\begin{equation}
\hat{Q}_i=e^\dagger_i e_i + d^\dagger_i d_i + \text{tr}(\textbf{p}^\dagger_i \textbf{p}_i) -1 = 0,
\end{equation}
and the equivalence between the fermion and boson representations  of the particle and spin densities
\begin{equation}
\hat{Q}^\mu_i =\text{tr} (\sigma^\eta \textbf{p}^\dagger_i \textbf{p}_i) +2\delta_{\mu,0} d^\dagger_i d_i -\sum_{\alpha\beta} f^\dagger_{i\alpha} \sigma^\mu_{\alpha\beta} f_{i\beta}=0.
\end{equation}
The saddle-point solution of the functional-integral for Eq. (\ref{Hsb}) corresponds to condensing all boson fields \{$e_i$, $d_i$, $p_{i\mu}$, $\lambda_i$, $\lambda^\mu_i$, $\chiup^\mu_{\langle ij\rangle}$\} and determining their values self-consistently by minimizing the state energy $\langle \widetilde{H} \rangle$.

\begin{figure*}
\begin{center}
\fig{7.in}{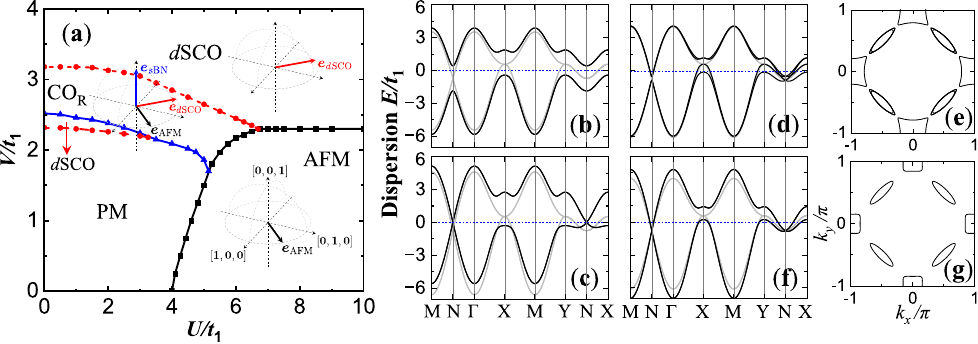}
\caption{(a) Ground state phase diagram of half-filled \tUVG\ model with fixed $\Gamma_2 = 0.2t_1$.
Solid and dashed lines denote, respectively, phase boundaries of first-order and continuous transitions.
Insets display the spin directions of order parameters, i.e., $\hat{\bm e}_\text{AFM}$, $\hat{\bm e}_{d\text{SCO}}$, and $\hat{\bm e}_\text{sBN}$, in each ground states.
Electronic structures of the nonparamagnetic ground states (black lines) along high-symmetry path at Coulomb interactions $(U, V)=$ $(6, 1.6)t_1$ for the insulating AFM (b), $(6, 3)t_1$ for the semimetallic $d$SCO (c), $(5, 2)t_1$ for metallic CO$_R$ (d), and $(1, 2.4)t_1$ for the metallic $d$SCO (f).
Grey lines denote the band dispersion of the PM states converged at the same parameters.
The corresponding Fermi surfaces of (d) and (f) are shown in, respectively, (e) and (g).}
\label{fig2}
\end{center}
\end{figure*}

The onsite and nn Coulomb repulsions, $U$ and $V$, would produce, respectively, magnetic and bond orders at sufficient strengths.
We consider two-sublattice solutions where onsite boson fields condense uniformly on each sublattice $\nu=A$ or $B$.
Explicitly, on site $i\in \nu$, $e_i=e_\nu$, $d_i=d_\nu$, $p_{i\mu}=p_{\nu \mu}$, $\lambda_i=\lambda_\nu$, and $\lambda^\mu_i =\lambda^\mu_\nu$.
Consequently, the local magnetic moment ${\bm m}_i$= ${\bm m}_\nu$ = ($m^x_\nu$, $m^y_\nu$, $m^z_\nu$), with component $m^\eta_\nu = \text{tr} (\sigma^\eta \textbf{p}^\dagger_\nu \textbf{p}_\nu)$, where a $g$-factor of 2 has been used.
Self-consistent calculations converge to charge uniform states with, if magnetism developed, perfect Ne\'el order in the local basis tracking the IrO$_6$ staggered rotation.
As a result, $e_\nu =e$, $d_\nu =d$, $\lambda_\nu =\lambda$, $p_{\nu 0} =p_0$, $\lambda^0_\nu = \lambda^0$, and $p_{A\eta} =-p_{B\eta} =p_\eta$, $\lambda^\eta_A = -\lambda^\eta_B =\lambda^\eta$ for $\eta=x,y,z$.
Consequently, the magnetic moment ${\bm m}_A = -{\bm m}_B = {\bm m}$.
Similarly, the condensation of boson fields on nn bonds can be expressed as a combination of $s$- and $d$-wave components
\begin{equation}
\chiup^\mu_{\langle ij\rangle} = \left( \chiup'_{\mu,s} +i\chiup''_{\mu,s} \right) + \tau_{ij} \left(\chiup'_{\mu,d} + i\chiup''_{\mu,d} \right), \quad i\in A \text{ and } j\in B,
\end{equation}
where $\chiup'_{\mu,s}$ ($\chiup''_{\mu,s}$) and $\chiup'_{\mu,d}$ ($\chiup''_{\mu,d}$) are the real and imaginary parts of the $s(d)$-wave component.
In the charge-uniform two-sublattice solutions, the contributions to the state energy per site from $V$ and $\Gamma_2$ are given by, respectively,
\begin{align}
E_V = & -V\sum_\mu \left( \chiup'^2_{\mu,s} + \chiup''^2_{\mu,s} +\chiup'^2_{\mu,d} + \chiup''^2_{\mu,d} \right), \label{eV} \\
E_{\Gamma_2} =& \Gamma_2 \left( \chiup'_{y,s} \chiup'_{y,d} + \chiup''_{y,s} \chiup''_{y,d} -\chiup'_{x,s} \chiup'_{x,d} - \chiup''_{x,s} \chiup''_{x,d} \right). \label{eG}
\end{align}
Clearly, in-plane anisotropy $\Gamma_2$ tends to bring mixture of $s$- and $d$-wave components in the $\mu=x, y$ bond orders to gain energy and, as a result, breaks the SO(3) rotational symmetry of spin currents down to $C_{4z}$ with four easy axes along [$\pm1$, $\pm1$, 0].
We note that nonzero $\chiup''_{0,s}$ brings charge currents flowing into or out of a lattice site from all the four connecting bonds, which violate the charge conservation and thus physically prohibited.
In contrast, nonzero ${\bm \chiup}_{s\text{SCO}} = (\chiup''_{x,s}, \chiup''_{y,s}, \chiup''_{z,s})$ $\equiv \chiup_{s\text{SCO}} \hat{\bm e}_{s\text{SCO}}$ that violates the spin conservation on each lattice site by generating $s$-wave spin current is however allowed at finite in-plane anisotropy $\Gamma_2$ since the presence of the latter requires the existence of spin-orbit coupling \cite{Jackeli2009prl}.
Similarly, the real parts of $d$-wave bond orders introduce charge bond nematicity (cBN) $\chiup_\text{cBN} =\chiup'_{0,d}$ \cite{Jiang2016prb} and sBN ${\bm \chiup}_\text{sBN} = (\chiup'_{x,d}, \chiup'_{y,d}, \chiup'_{z,d}) \equiv \chiup_\text{sBN} \hat{\bm e}_\text{sBN}$, while the imaginary parts of $d$-wave bond orders generate SFP $\chiup_\text{SFP} =\chiup''_{0,d}$ and $d$SCO with ${\bm \chiup}_{d\text{SCO}} = (\chiup''_{x,d}, \chiup''_{y,d}, \chiup''_{z,d}) \equiv \chiup_{d\text{SCO}} \hat{\bm e}_{d\text{SCO}}$ for spins along $\hat{\bm e}_{d\text{SCO}}$ direction.

To this end, in order to obtain all the possible states, we use different initial conditions for solving the self-consistency equations numerically.
Besides PM, AFM, and $d$SCO, we also encounter converged metallic solutions of coexisting AFM and $d$SCO, which induces additional sBN and brings about interesting chirality for the spin directions of these three orders.
When more than one converged states exist at a given set of parameters, we compare their energies to determine the true ground state.

\section{III. Results and Discussions} \label{results}

\textit{Phase diagram at half filling}.
We first explore the zero-temperature phase structure and the formation of $d$SCO in the half-filled $t$-$U$-$V$-$\Gamma_2$ model for a given in-plane anisotropy.
At $\Gamma_2=0.2t_1$, the ground state phase diagram is presented in Fig. \ref{fig2}a in the plane spanned by the on-site Coulomb repulsion $U$ and nn Coulomb repulsion $V$.
It consists of PM, AFM, $d$SCO, and CO$_R$.
Here, CO$_R$ refers to the coexisting state of AFM, $d$SCO, and sBN, with the subscript denotes the right-handed chirality formed by the spin directions of these three orders.
The solid and dashed lines denote, respectively, a first-order and a continuous phase transition between two neighboring phases, and the boundaries are determined by comparing the state energies of different phases.

\textit{PM and AFM}.
In the regime near the phase diagram origin where neither $U$ nor $V$ is strong enough to produce magnetic or bond orders, the ground state is a PM metal with the bandwidth renormalized by electron correlations.
Increasing $U$ at small $V$, the PM metal gives way to the AFM insulator at a critical $U$ via a first-order transition.
As shown in Fig. \ref{fig2}a, the critical $U$ increases as one enhances the nn Coulomb repulsion $V$, since the latter effectively enlarges the bandwidth.
The AFM gaps out the DP and QBT simultaneously, resulting to a band dispersion displayed in Fig. \ref{fig2}b at $(U, V)=(6, 1.6)t_1$.
Since $\Gamma_2$ does not generate any in-plane magnetic anisotropy on the mean-field level \cite{Jackeli2009prl} and all other terms in the \tUVG\ model are invariant under spin rotation, the moment direction of AFM obtained here thus has the SO(3) rotation symmetry.
Inclusion of out-of-plane anisotropy $\Gamma_1$, which is shown to be present in \sr\ \cite{Jackeli2009prl}, would break SO(3) symmetry down to SO(2) with easy $xy$-plane.
Furthermore, when coexisting with $d$SCO, as we shall show later, the spin rotation symmetry is further reduced to $C_{4z}$ with four equivalent easy-axes along [$\pm 1$, $\pm$ 1, 0], consistent with experimental observations and theoretical results in more sophisticated five-orbital models~\cite{Boseggia2013condmatt,Crawford1994prb,YeFeng2013prb,HongYunjeong2016prb,Nauman2017prb,Nauman2022conmat,Zhou2018theophy,Mohapatra2021condmatt,Khaliullin2019prl}.
In this work, we neglect $\Gamma_1$ in the Hamiltonian for simplicity and, without loss of generality, fix the moment direction along [1, 1, 0], unless otherwise noted.

Interestingly, we note that AFM could not be stabilized by any $U$ when nn Coulomb repulsion is large, i.e., $V \gtrsim 2.3t_1$, as shown in the phase diagram Fig. \ref{fig2}a.
In the strongly coupling theory, the dynamically generated AFM Zeeman field scales with the average kinetic energy instead of $U$ \cite{ZhouSen2010prl}.
This effectively sets a upper limit on the energy gain via the formation of AFM.
On the other hand, the energy gain of bond order, if developed, is on the order of $V$, as shown clearly in Eq. (\ref{eV}).
Therefor, AFM is unable to compete with bond orders, $d$SCO in particular here, when $V$ is sufficiently large, regardless of how strong on-site $U$ is.

\textit{dSCO}. When $V$ is large, it drives charge or spin currents on the nn bonds and consequently leads to SFP or $d$SCO separately.
These two states are energetically degenerate and share the identical band structure in the absence of $\Gamma_2$, and $d$SCO has SO(3) rotation symmetry associated with the spin direction of the spin current.
The SFP breaks TRS while $d$SCO does not, and thus these two states can never coexist with each other.
We note that the in-plane anisotropy $\Gamma_2$ is able to lift the degeneracy within the mean-field theory and stabilize $d$SCO with circulating spin currents as the ground state.
Furthermore, in the presence of $\Gamma_2$, the SO(3) rotation symmetry of $d$SCO is broken down to $C_{4z}$ with four equivalent preferred directions along [$\pm1$, $\pm1$, 0], i.e., ${\bm \chiup}_{d\text{SCO}}=(\pm1, \pm1, 0)\chiup''_d$.
The lifting of degeneracy and the lowering of rotation symmetry are achieved via the development of a $s$-wave spin current order ($s$SCO) with the spin direction along [$\pm1$, $\mp1$, 0] direction, i.e., ${\bm \chiup}_{s\text{SCO}}=(\pm1, \mp1, 0)\chiup''_s$, which coexists with $d$-wave ${\bm \chiup}_{d\text{SCO}}$ and lowers the state energy per site in Eq. (\ref{eG}) by $-2\Gamma_2 \chiup''_s \chiup''_d$.
Due to its $d$-wave nature, $d$SCO gaps out the QBT at $X$ but leaves the DP at $N$ unaltered.
At sufficient strong $V$ where the $d$SCO gap is large, it gives rise to a Dirac semimetal with vanishing density of states at Fermi level, as illustrated in the band dispersion at $(U, V) =(6, 3)t_1$ shown in Fig. \ref{fig2}c.

\begin{figure}
\begin{center}
\fig{3.4in}{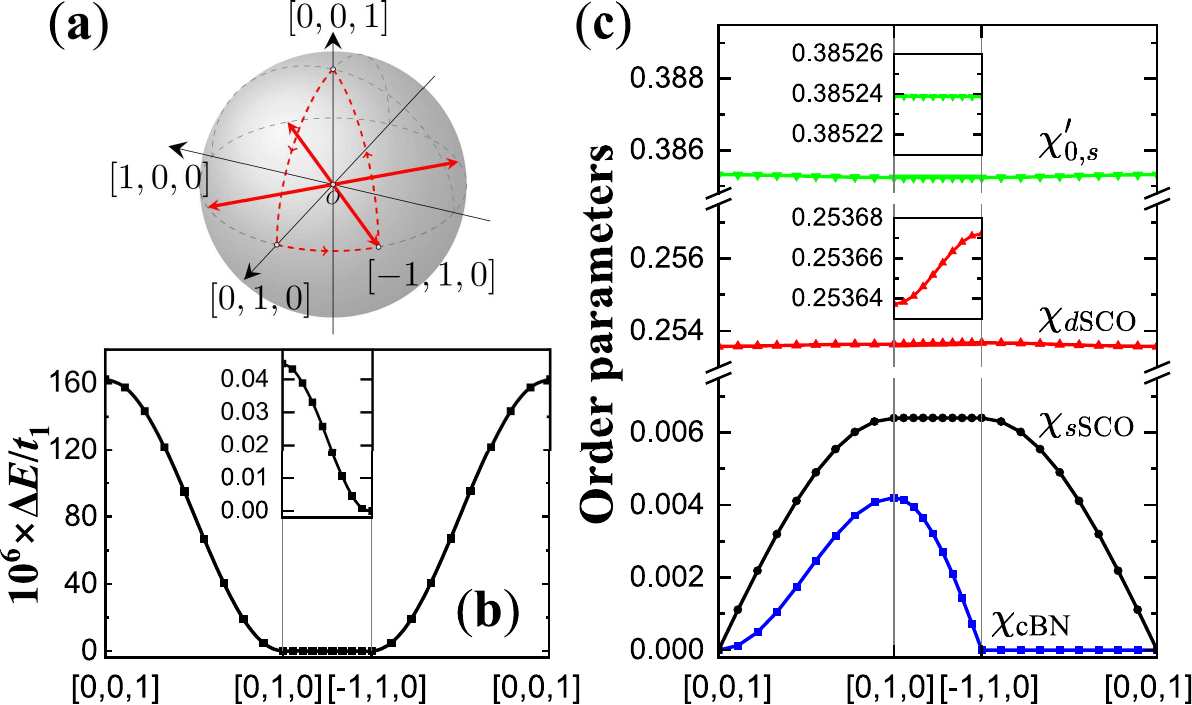}
\caption{(a) Path of $\hat{\bm e}_{d\text{SCO}}$ for the variational calculations. (b) The state energy and (c) the magnitudes of various bond orders in $d$SCO as a function of the pinned spin direction $\hat{\bm e}_{d\text{SCO}}$ displayed in (a).
The Coulomb interactions $(U, V) =(6, 3)t_1$.}
\label{fig3}
\end{center}
\end{figure}

In order to elaborate on the anisotropy in the spin direction of $d$SCO in the presence of $\Gamma_2$, we have performed variational calculation where an external field is applied to pin the spin direction of spin currents along the desired direction, whereas the elastic part of the external field is excluded in the state energy.
As a function of $\hat{\bm e}_{d\text{SCO}}$ pinned along the path shown in Fig. \ref{fig3}a, the state energy per site is plotted in Fig. \ref{fig3}b at $(U, V) = (6, 3) t_1$, with the inset focus on the in-plane anisotropy.
For clarity, all energies are shown with respect to $E_{[-1, 1, 0]}$, i.e., energy of $d$SCO with $\hat{\bm e}_{d\text{SCO}}$ along $[-1, 1, 0]$.
Clearly, $[-1, 1, 0]$ is the most preferred spin direction for $d$SCO, with a tiny in-plane anisotropy $E_{[0,1,0]}-E_{[-1, 1, 0]} \simeq 4.5\times 10^{-8} t_1$ and a relatively large out-of-plane anisotropy $E_{[0,0,1]}-E_{[-1, 1, 0]} \simeq 1.6\times 10^{-4} t_1$.
The evolution of the amplitudes of various bond orders are displayed in Fig. \ref{fig3}c.
It is clear that the in-plane $d$SCO gain energy via the formation of ${\bm \chi}_{s\text{SCO}}$, while the [-1, 1, 0] $d$SCO lowers energy further by the small increasing in its amplitude $\chiup_{d\text{SCO}}$.

\textit{CO$_R$}.
Reducing the nn $V$ from the $d$SCO phase at small to moderate $U$, the phase diagram in Fig. \ref{fig2}a shows a wide regime where the ground state possesses simultaneously AFM, $d$SCO, and sBN orderings.
Interestingly, the spin directions of these three orders, $\hat{\bm e}_\text{AFM}$, $\hat{\bm e}_{d\text{SCO}}$, and $\hat{\bm e}_\text{sBN}$, are perpendicular to each other and form a right-handed chirality, i.e., $(\hat{\bm e}_\text{AFM} \times \hat{\bm e}_{d\text{SCO}}) \cdot \hat{\bm e}_\text{sBN} =1$, as displayed in the inset in Fig. \ref{fig2}a.
Therefore, we refer to this coexistent state as CO$_R$ hereafter.
At $(U, V)=(5, 2)t_1$, the converged CO$_R$ has AFM moment ${\bm m}= 0.135 \mu_B$, $d$SCO $\chiup_{d\text{SCO}} =0.117$, and sBN $\chiup_\text{sBN} =0.044$, with the electronic structure shown in Fig. \ref{fig2}d.
The primary feature of the coexistence lies at $X$ point where the doubly degenerate bands in AFM and $d$SCO are now split unevenly, unnoticeably small on the conduction bands above the Fermi level but substantially larger on the valence bands crossing the Fermi level.
It gains energy by pushing one of the valence band below Fermi level.
The corresponding Fermi surfaces is displayed in Fig. \ref{fig2}e, exhibiting two elliptic electron pockets around $N$ points and a square-shaped hole pocket around $X$ and $Y$ points.

\begin{figure}
\begin{center}
\fig{3.4in}{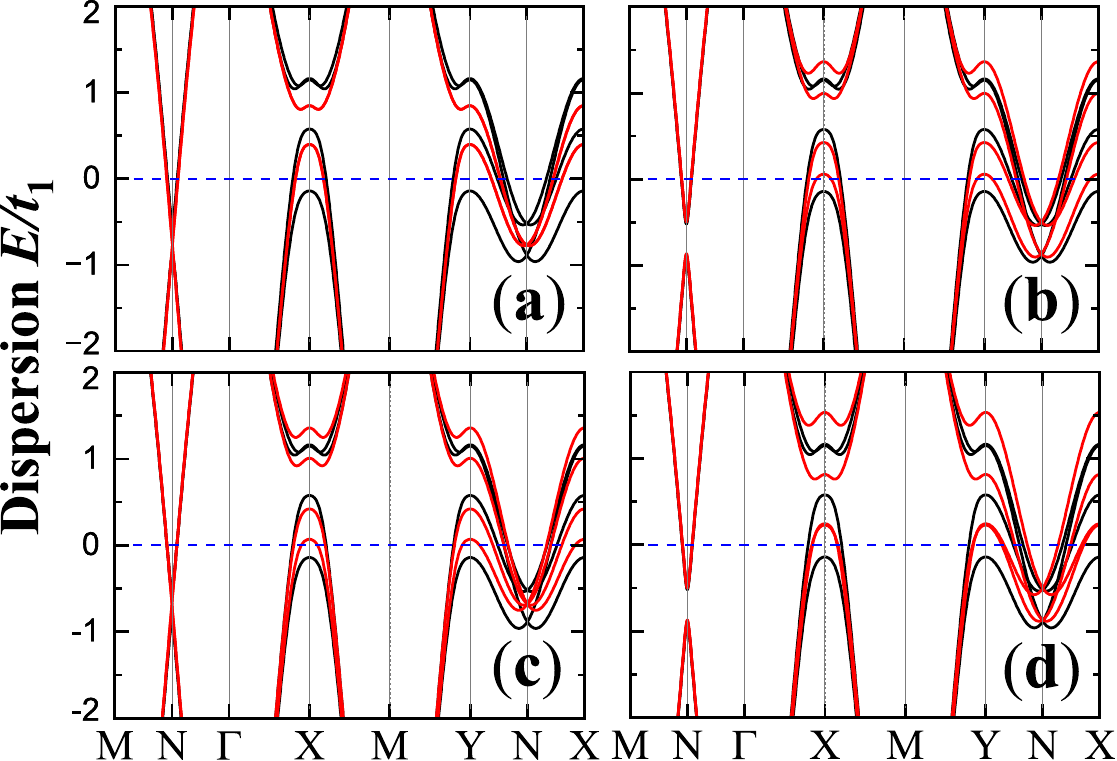}
\caption{Low-energy band dispersions (red lines) of (a) the self-consistently converged pure $d$SCO, (b) CO$_R$ with sBN order manually switched off, (c) CO$_R$ with AFM order manually switched off, and (d) CO$_R$ with the spin direction of sBN manually reversed.
Band dispersions of the self-consistently converged CO$_R$ are plotted in black lines for comparison.
The Coulomb interactions $(U, V)=(5, 2)t_1$.}
\label{fig4}
\end{center}
\end{figure}

To shed light on the emergence of the coexisting CO$_R$ phase, we plot in Fig. \ref{fig4}a the band structure of the would-be pure $d$SCO self-consistently converged at the same interactions, $(U, V)=(5, 2)t_1$.
Clearly, the $d$SCO gap is not large enough to realize a Dirac semimetal, as the doubly degenerate valence bands cross the Fermi level around $X$ point.
This would leads to a large density of states at Fermi level, which is expected to trigger the emergence of AFM and sBN, and hence the stabilization of the coexistent state.
To further understand the coexistence and its chirality, we manually switch off the sBN or AFM in the CO$_R$ and plot the resulting electronic structure in, respectively, Fig. \ref{fig4}b and \ref{fig4}c.
Clearly, in both cases, the conduction and valence bands split evenly at $X$ point, failing to push one of the valence band below Fermi level and thus could not compete with CO$_R$ in lowering state energy.
Furthermore, symmetry analysis conducted in Appendix B shows that combination of any two of $d$SCO, AFM, and sBN has the same symmetry properties as the coexisting state of all these three orders.
This implies that the combination of any two of these three orders would in principal give rise to the emergence of the third order, which has been verified numerically in Appendix B.
This is exactly the reason why the regime of CO$_R$ extends to $U=0$ (this is the case even in the absence of $\Gamma_2$) in the phase diagram Fig. \ref{fig2}a, where kinetic magnetism arises in the absence of local repulsion.
Next, we manually switch the chirality from right-handed to left-handed, which can be achieved by reversing the spin direction of all three orders or only one of them.
The resulting band dispersions are all identical and shown in Fig. \ref{fig4}d.
Interestingly, the splitting in the conduction bands is now much larger than that in the valence bands, which is disadvantageous for lowering energy and thus unstable.

We note that, underneath the regime of CO$_R$ in the phase diagram Fig. \ref{fig2}a, there is another small regime for pure $d$SCO.
The band dispersions and corresponding Fermi surfaces are shown in Figs. \ref{fig2}f and \ref{fig2}g at $(U, V) = (1, 2.4)t_1$.
The coexistent state could not be stabilized in this regime, as we shall show later, because it fails to push one of the valence band below Fermi level at $X$ point to lower energy.

\begin{figure}[t!]
\begin{center}
\fig{3.4in}{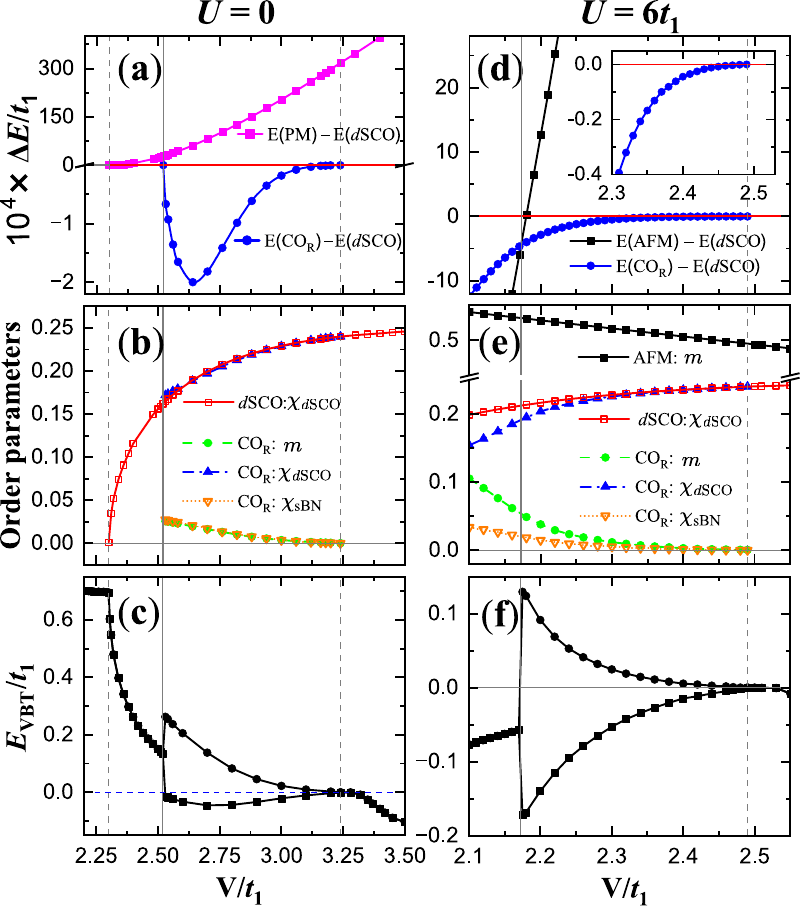}
\caption{(a,d) The state energy per site with respect to $d$SCO, (b,e) the magnitudes of magnetism and bond orders, and (c,f) the locations of valence bond tops at $X$ point, $E_\text{VBT}$, as a function of the nn Coulomb repulsion $V$.
The onsite Coulomb repulsion $U=0$ in left panels and $U=6t_1$ in right panels.}
\label{fig5}
\end{center}
\end{figure}

\textit{Phase transitions at half filling.}
To investigate in detail the phase transitions between different ground states shown in Fig. \ref{fig2}a, we fix onsite Coulomb repulsion $U = 0$ or $6t_1$ and monitor the phase evolution as a function of nn Coulomb repulsion $V$, focusing on the regions near phase boundaries.
The state energies per site with respect to $d$SCO are compared in Fig. \ref{fig5}a and \ref{fig5}d.
The magnitudes of magnetism and bond orders are shown in Fig. \ref{fig5}b and \ref{fig5}e, and the locations of valence band top $E_\text{VBT}$, i.e., the energies of the lower two levels at $X$ point, are plotted in Fig. \ref{fig5}c and \ref{fig5}f.

For $U=0$, self-consistent calculations at $V\gtrsim 3.24 t_1$ can converge to PM and $d$SCO, with the latter being the ground state with much lower energy, as shown in Fig. \ref{fig5}a.
The magnitude of $d$SCO order is quite large, $\chiup_{d\text{SCO}} \gtrsim 0.24$ (Fig. \ref{fig5}b), which gaps the QBT by a large gap and consequently leads to a Dirac semimetal shown in Fig. \ref{fig2}c, with the two valence band tops (Fig. \ref{fig5}c) degenerate in energy and lying below the Fermi level.
As one reduces nn $V$, $E_\text{VBT}$ increase in energy and cross the Fermi level at $V\simeq 3.24 t_1$, where the $d$SCO gives way to CO$_R$ via a second-order transition.
AFM and sBN orders start to develop, and their coexistence with $d$SCO split the two valence bands and push one of them below the Fermi level to lower energy.
As shown in Fig. \ref{fig5}c, the evolution of the lower valence band top is nonmonotonic and it is about to cross the Fermi level again at $V\simeq 2.51 t_1$.
Here the ground state changes from CO$_R$ to $d$SCO via a first-order transition, as indicated by the kink in state energy (Fig. \ref{fig5}a) and the abrupt disappearance of AFM and sBN orders (Fig. \ref{fig5}b).
Reducing $V$ further, the $d$SCO is suppressed gradually and vanishes at $V\simeq 2.30t_1$ where the ground state becomes PM via a second-order phase transition.

For fixed onsite Coulomb repulsion $U=6t_1$, the ground state at $V\gtrsim 2.49t_1$ is also a Dirac semimetal, i.e., $d$SCO with large $d$-wave gap.
Reducing nn $V$, the ground state first undergoes a second-order transition from $d$SCO to CO$_R$ at $V\simeq 2.49t_1$ where the doubly degenerate valence band top reaches Fermi level, as shown in Fig. \ref{fig5}f.
At $V\simeq 2.17t_1$, AFM sets in via a first-order transition, as shown by the level crossing displayed in Fig. \ref{fig5}d.

\begin{figure}[t!]
\begin{center}
\fig{3.4in}{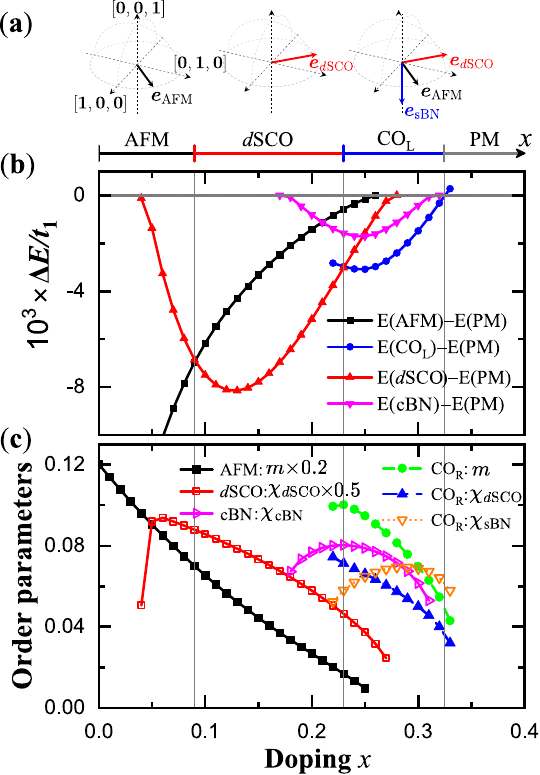}
\caption{(a) Zero-temperature phase diagram as a function of electron doping, with the insets show the spin directions of AFM, $d$SCO, and sBN orders.
(b) Energy per site of converged states with respect to PM state. (c) Magnitudes of the magnetic and bond orders in nonparamagnetic phases.
Here the in-plane anisotropy $\Gamma_2=0.2 t_1$ and the Coulomb repulsions $(U, V)=(6, 1.6)t_1$.}
\label{fig6}
\end{center}
\end{figure}

\textit{Electron-doping evolution of AFM insulator.}
To make connection to the experimental observations in electron-doped \sr~\cite{Torre2015prl, Kim2014sic, FengDL2015prx, Kim2016natphy, Chenxiang2015prb}, it is instructive to study the state evolution with electron doping away from half filling.
We fix the in-plane anisotropy $\Gamma_2 = 0.2t_1$ and the Coulomb repulsions $(U, V) = (6, 1.6)t_1$, where the ground state at stoichiometry is the insulating AFM, as shown in Fig. \ref{fig2}a.
As a function of electron doping $x$ away from the half filling, the ground state undergoes successively phase transitions from the AFM to $d$SCO at $x\simeq 0.09$,  to a coexistent state with left-handed chirality (CO$_L$) at $x\simeq 0.23$, and finally to PM at $x \simeq 0.32$, as displayed in Fig. \ref{fig6}a.
The energies per site of converged states at different electron doping are summarized in Fig. \ref{fig6}b, with respect to the energy of the PM phase.
In addition to the four ground states shown in Fig. \ref{fig6}a, we managed to converge to the cBN state in the electron doping range $0.18 \leq x \leq 0.31$, but it is always higher in energy and fails to present itself as a ground state.
The features of level-crossing between these four ground states in Fig. \ref{fig6}b indicate that the three successive phase transitions are all of first-order.
Fig. \ref{fig6}c shows the amplitudes of AFM moment and bond orders in the nonparamagnetic phases.

\begin{figure}[t!]
\begin{center}
\fig{3.4in}{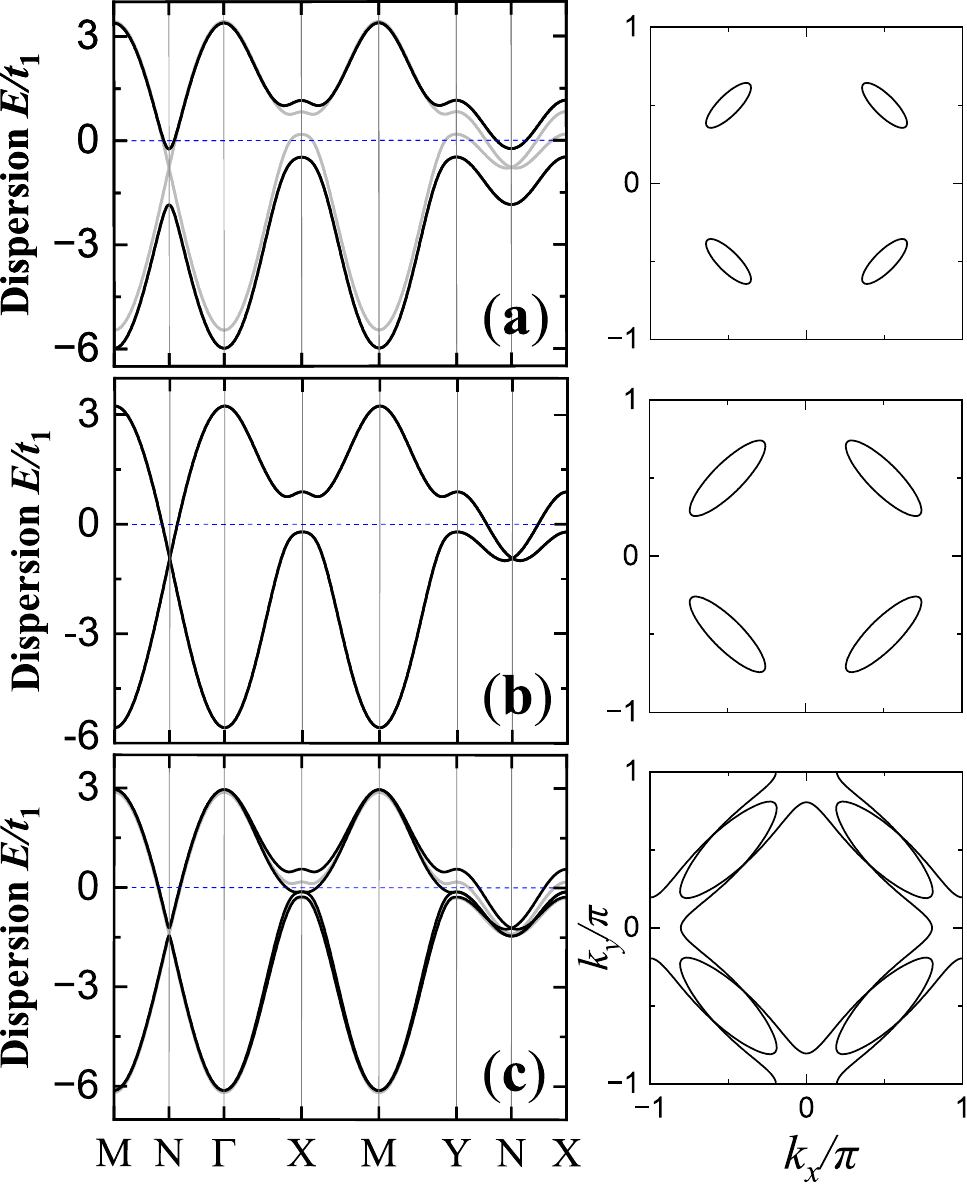}
\caption{Band dispersions (black lines) and corresponding Fermi surfaces of (a) AFM at $x=0.04$, (b) $d$SCO at $x=0.1$, and (c) CO$_L$ at $x=0.25$.
The grey lines in (a) and (c) denote the electronic structure of $d$SCO converged at the corresponding electron doping.}
\label{fig7}
\end{center}
\end{figure}

The band dispersions and corresponding Fermi surfaces of AFM at $x=0.04$, $d$SCO at $x=0.1$, and CO$_L$ at $x=0.25$ are plotted in Fig. \ref{fig7}.
At half filling $x=0$, the ground state is an AFM insulator.
In the $x=0.04$ electron-doped AFM, additional electrons go to the bottom of conduction bands around $N$ point and give rise to the elliptic electron pockets showing in Fig. \ref{fig7}a.
The $d$SCO away from half filling in Fig. \ref{fig7}b is a doped Dirac semimetal as the $d$-wave order gaps only the QBT but leave the Dirac point unaltered.
Increasing electron doping $x$ further, the chemical potential moves upwards to accommodate more electrons, and the $d$SCO order decreases in amplitude as well.
As a result, the bottom of the conduction bands near $X$ point get closer to the Fermi level.
However, the coexistent state sets in as the ground state before the bottom of the conduction bands reach the Fermi level.
The coexistence split the doubly degenerate conduction band bottoms at $X$ point and push one of them below Fermi level to lower energy, as shown in Fig. \ref{fig7}c.
Interestingly, this coexistent state tends to split more the conduction band, thus its chirality is left-handed with $(\hat{\bm e}_\text{AFM} \times \hat{\bm e}_{d\text{SCO}}) \cdot \hat{\bm e}_\text{sBN} =-1$, instead of right-handed for the coexistent state at half-filling.

\textit{Manipulation of chirality by carrier doping.}
We have shown that when the valence band top or conduction band bottom near $X$ point is above and close to the Fermi level in $d$SCO phase, coexistent state can be stabilized by splitting the corresponding bands, pushing one of them below Fermi level, and thus lowering the ground state energy.
Furthermore, the chirality of the coexistent state manifests itself in the splitting of the bands, CO$_R$ splits mainly the valence bands while CO$_L$ splits more the conduction bands, as shown clearly in Figs. \ref{fig2}d and \ref{fig7}c.
It thus suggests that the emergence of coexistent state and its chirality can potentially be manipulated by gate-voltage or carrier doping.
To demonstrate it, we start with a Dirac semimetal at half filling by setting $(U, V) = (6, 3) t_1$, where the large $d$SCO order produces a wide energy gap near $X$ point, as shown in Fig. \ref{fig2}c.
The resulting valence band top is below Fermi level and the conduction band bottom is quite far away from the Fermi level, neither of them can trigger the development of coexistent state and the ground state is thus a Dirac semimetal with pure $d$SCO.
Electron- and hole-doping would move the Fermi level, respectively, upwards and downwards, and $d$SCO is expected to give its way to coexistent state with different chirality at critical doping concentrations.

\begin{figure}[t!]
\begin{center}
\fig{3.4in}{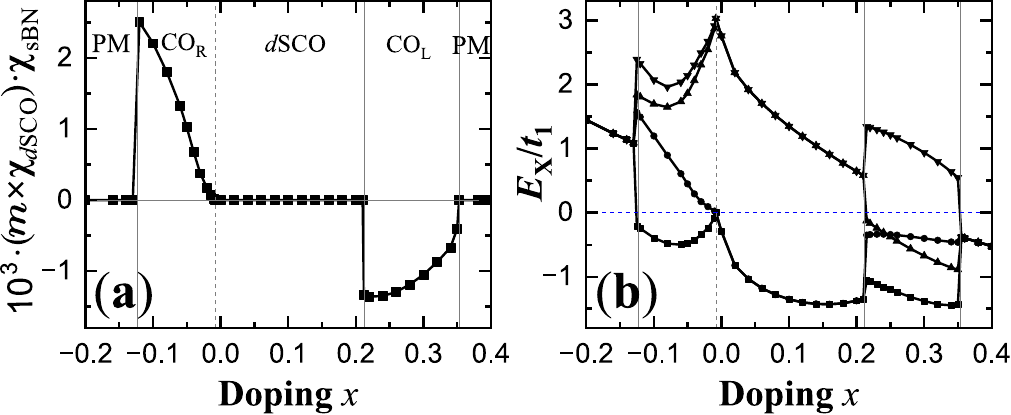}
\caption{Doping dependence of (a) the ground state chirality $({\bm m} \times {\bm \chiup}_{d\text{SCO}}) \cdot {\bm \chiup}_\text{sBN}$ and (b) the four energy levels at $X$ point.
The Coulomb repulsions $(U, V) = (6, 3)t_1$.}
\label{fig8}
\end{center}
\end{figure}

The state evolution of the Dirac semimetal is summarized in Fig. \ref{fig8} as a function of carrier doping $x$, with positive $x$ stands for electron doping and negative $x$ for hole doping.
The doping dependence of the ground state chirality $({\bm m} \times {\bm \chiup}_{d\text{SCO}}) \cdot {\bm \chiup}_\text{sBN}$  is displayed in Fig. \ref{fig8}a.
It obtains a nonzero value only in the coexistent state, positive in CO$_R$ with right-handed chirality and negative in CO$_L$ with left-handed chirality.
Clearly, CO$_R$ and CO$_L$ can be realized separately by hole- and electron-doping a Dirac semimetal.
Fig. \ref{fig8}b shows the doping evolution of the four energy levels at $X$ point.
As electrons doped into the Dirac semimetal, the Fermi level gets closer to the conduction band bottom, i.e, the upper two levels at $X$ point, and $d$SCO gives its way to CO$_L$ via a first-order transition at critical doping $x\simeq 0.21$ where the coexistence splits mainly the conduction bands and pushes one of the conduction band bottom below Fermi level.
The CO$_L$ becomes unstable with respect to PM at $x\gtrsim 0.35$.
On the other hand, hole doping drives the Fermi level of the doped Dirac semimetal towards the valence band top, and they coincide in energy at $x\simeq -0.01$, where CO$_R$ sets in via a second-order transition.
The coexistence splits mainly the valence bands and pushes one of the valence band top below Fermi level to lower the ground state energy.

\section{IV. Summaries}
In this paper, we constructed an effective square lattice single-orbital \tUVG\ model for the low-energy physics in \sr, where $\Gamma_2$ is the in-plane anisotropy of the $J_\text{eff}=\frac{1}{2}$ pseudospins arising from the cooperative interplay between Hund's rule coupling and SOC~\cite{Jackeli2009prl}.
To study the mean-field ground state properties of the model,
the onsite Coulomb repulsion $U$ is treated by SU(2) spin-rotation invariant slave-boson mean-field theory, while nn interactions, $V$ and $\Gamma_2$, are mean-field decoupled into bond channels.
We obtain the ground state phase diagram at half-filling with fixed $\Gamma_2=0.2 t_1$, in the plane spanned by $U$ and $V$, and investigate the competition and coexistence of magnetic and bond orders.

The nn $V$, which is expected to be significant in \sr\ due to the large spatial extension of the Ir 5$d$ orbitals, is capable of driving $d$SCO with circulating spin current, which is degenerate in energy with the SFP produced by circulating charge current.
We demonstrated the importance of in-plane anisotropy $\Gamma_2$ in the stabilization of $d$SCO.
It lifts the degeneracy by stabilizing $d$SCO and lowers the rotation symmetry of $d$SCO from SO(3) to $C_{4z}$ with four equivalent preferred directions along [$\pm 1$, $\pm 1$, 0].
The lifting of degeneracy and the lowering of rotation symmetry are achieved via the development of a $s$-wave spin current with the spin direction along [$\pm1$, $\mp1$, 0].
It coexists with $d$-wave ${\bm \chiup}_{d\text{SCO}}$ and lowers the state energy.
The $d$SCO order gaps out the QBT while leaves the DP unaltered, giving rise to a Dirac semimetal when its amplitude is sufficiently large.
Thus, we have succeeded in providing a minimal effective single-orbital model for the hidden electronic order capable of describing the unexpected breaking of spatial symmetries in the AFM ordered spin-orbit Mott insulator in stoichiometric \sr\ and the unconventional pseudogap phenomena in electron doped \sr~\cite{ZhouSen2017prx}.

Remarkably, we discovered that, in a wide regime of the phase diagram, the ground state is CO$_R$, a coexisting phase of AFM, $d$SCO, and sBN with unconventional right-handed chirality, i.e., $(\hat{\bm e}_\text{AFM} \times \hat{\bm e}_{d\text{SCO}}) \cdot \hat{\bm e}_\text{sBN} =1$.
In this phase, the $d$SCO order is relatively small, and the Fermi level would cross the valence band near $X$ point in the would-be pure $d$SCO state.
The coexistent state sets in here as the ground state, gaining energy by splitting the bands and pushing one of the valence band below Fermi level.
Its chirality leads to uneven splitting of the valence and conduction bands, with CO$_R$ splits mostly the valence band and CO$_L$ splits mainly the conduction bands.
We demonstrated that the emergence of coexistent state and its chirality can potentially be manipulated by carrier doping a Dirac semimetal.
CO$_R$ with right-handed chirality and CO$_L$ with left-handed chirality can be realized separately by hole and electron doping.
The electron doping evolution of the AFM insulator is also investigated.
As increases the electron doping concentration $x$, the ground state undergoes successively first-order phase transitions from the AFM to $d$SCO, CO$_L$, and PM.

It would be interesting to investigate if the coexistent phase possesses any nontrival topological properties.
In the mean-field study of the \tUVG\ model conducted in this work, the coexistent states, either CO$_R$ or CO$_L$, can emerge only in metallic states, no insulating states with simultaneously developed AFM and $d$SCO orders have been stabilized.
It would be desirable to study the quantum states and phase diagram of this model by more sophisticated analytical and numerical methods for an improved understanding.
The findings presented in this work can be viewed as a starting point for further studies.

\section{Acknowledgments}

JD and SZ are supported by the National Key R\&D Program of China (Grant No. 2022YFA1403800), the Strategic Priority Research Program of CAS (Grant No. XDB28000000) and the National Natural Science Foundation of China (Grants Nos. 11974362, 12047503, and 12374153).
ZW is supported by the U.S. Department of Energy, Basic Energy Sciences (Grant No. DE-FG02-99ER45747) and by the Research Corporation for Science Advancement (Cottrell SEED Award No. 27856).
Numerical calculations in this work were performed on the HPC Cluster of ITP-CAS.

\appendix

\begin{figure}[h!]
\begin{center}
\fig{2.8in}{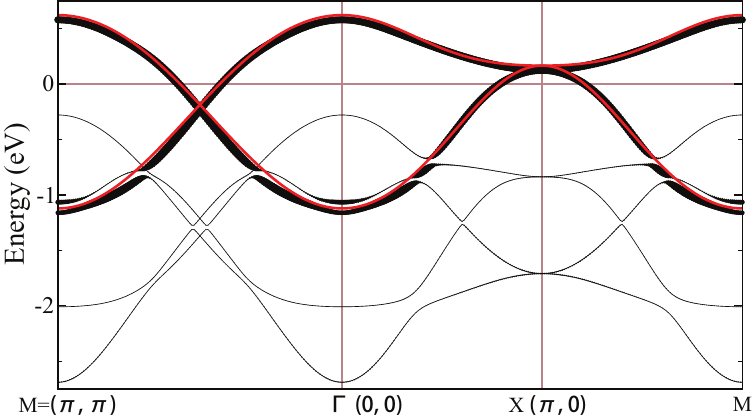}
\caption{Band dispersion (red lines) of the single-orbital tight-binding model.
The black lines display the interacting electronic structure of undoped \sr\ obtained from the five-orbital model \cite{ZhouSen2017prx} at Hubbard interaction $(U, J_H)=(1.2, 0.05)$ eV, with the line thickness denoting the content of $J_\text{eff}=1/2$ doublet.}
\label{fig9}
\end{center}
\end{figure}

\section{Appendix A: Single-orbital tight-binding model} \label{app1}
In the atomic limit, Ir$^{4+}$ has a 5$d^5$ configuration, with 5 electrons occupy the lower threefold $t_{2g}$ orbitals separated from the higher twofold $e_g$ orbitals by the cubic crystal field.
The strong atomic SOC splits the $t_{2g}$ orbitals into a low-lying $J_\text{eff}=\frac{3}{2}$ multiplet occupied by 4 electrons and a singly occupied $J_\text{eff}=\frac{1}{2}$ doublet.
For \sr\ solid, electronic structure calculations using the local-density approximation including SOC and structural distortion shows that, for the realistic bandwidths and crystalline electric field, the atomic SOC is insufficient to prevent two bands of predominantly $J_\text{eff} =\frac{1}{2}$ and $\frac{3}{2}$ characters to cross the Fermi level and give rise to two Fermi surfaces.
It has been shown that electron correlations, i.e., multiorbital Hubbard interaction, leads to a significantly enhanced effective SOC for the $t_{2g}$ complex~\cite{Pesin2010natphy, ZhouSen2017prx, JiangKun2023cpl}, pushing the $J_\text{eff} =\frac{3}{2}$ band in the LDA band structure below the Fermi level, and consequently gives rise to the single band crossing the Fermi level that is of dominant $J_\text{eff} =\frac{1}{2}$ character~\cite{Kim2008prl}.
This correlation-induced band polarization through enhancement of the SOC by the Hubbard interaction enables the single-orbital model of $J_\text{eff} =\frac{1}{2}$ electrons for \sr.

Fig. \ref{fig9} displays the interacting electronic structure (black lines) of undoped \sr\ obtained from the realistic five-orbital model \cite{ZhouSen2017prx} at Hubbard interaction $(U, J_H)=(1.2, 0.05)$ eV, with the line thickness denotes the content of the $J_\text{eff} =\frac{1}{2}$ doublet.
The red lines show the single-orbital tight-binding dispersion with hopping parameters chosen according to $t_{ij}=(218, 52, -18)$ meV for the first, second, and third nearest neighbors, respectively.
Clearly, the effective single-orbital model describes faithfully the low-energy electronic structure of the interacting \sr.

\begin{table}[htb]
\hspace{-2.7cm}
\begin{ruledtabular}
\begin{tabular}{c|cccccccc}
 & $E$ & $C_{4z}$ & $C_{2z}$ & $C^{-1}_{4z}$ & $M_x$ & $M_y$ & $M_{xy}$ & $M_{-xy}$ \\
\hline
AFM & 0 & $\times$ & \ttai & $\times$ & $\times$ & $\times$ & \ttai & 0  \\
$d$SCO & 0 &  $\times$ & \ttai & $\times$ & $\times$ & $\times$ & \ttai & 0 \\
sBN & 0 & $\times$ & 0, \ttai & $\times$ & $\times$ & $\times$ & 0, \ttai & 0, \ttai  \\
\hline
\hline
& $T$ & $TC_{4z}$ & $TC_{2z}$ & $TC^{-1}_{4z}$ & $TM_x$ & $TM_y$ & $TM_{xy}$ & $TM_{-xy}$ \\
\hline
AFM & \ttai & $\times$ & 0 & $\times$ & $\times$ & $\times$ & 0 & \ttai \\
$d$SCO & 0 & $\times$ & \ttai & $\times$ & $\times$ & $\times$ & \ttai & 0 \\
sBN & $\times$ & 0, \ttai & $\times$ & 0, \ttai & 0, \ttai & 0, \ttai & $\times$ & $\times$
\end{tabular}
\end{ruledtabular}
\caption{Symmetries of [1, 1, 0]-ordered AFM, [$-1$, 1, 0]-ordered $d$SCO, and [0, 0, 1]-ordered sBN states.
The table gives the lattice translation required for a state to recover itself after a symmetry operation of the magnetic space group of $4mm1'$.
\ttai\ is the translation of one lattice constant along either $x$- or $y$-axis.
Symbol $\times$ means such a lattice translation does not exist.} \label{symm}
\end{table}

\section{Appendix B: Symmetry analysis}\label{app2}
We analyze the symmetry properties of the states considered in this work within the symmetry point group $C_{4v}\otimes \{E, T\} $, i.e., the magnetic point group $4mm1'$, where $C_{4v}$ is the symmetry point group of the two-dimensional square lattice and $T$ is the time-reversal operation.
The magnetic point group $4mm1'$ has 16 symmetry operations: the 8 symmetry operations of $C_{4v}$ listed in the top half of Table \ref{symm} and their products with $T$ in the bottom half.
The 8 operations are identity $E$, fourfold rotation around $z$-axis $C_{4z}$, twofold rotation around $z$-axis $C_{2z}$, inverse fourfold rotation around $z$-axis $C^{-1}_{4z}$, mirror reflection about $yz$-plane $M_x$, mirror reflection about $zx$-plane $M_y$, mirror reflection about diagonal-plane $M_{xy}$, and mirror reflection about anti-diagonal-plane $M_{-xy}$.
The symmetries of [1, 1, 0]-ordered AFM, [$-1$, 1, 0]-ordered $d$SCO, and [0, 0, 1]-ordered sBN are summarized in Table \ref{symm}, which gives the lattice translation, if it exists, required for a state to recover itself after a symmetry operation of the magnetic point group $4mm1'$.
The state does not have the corresponding symmetry if it could not recover itself by any lattice translation after a symmetry operation.
Clearly, both [1, 1, 0]-ordered AFM and [$-1$, 1, 0]-ordered $d$SCO break \{$C_{4z}$,  $C^{-1}_{4z}$, $M_x$, $M_y$, $TC_{4z}$, ${TC^{-1}_{4z}}$, $TM_x$, $TM_y$\} symmetries and belong to the magnetic point group $mm21'$, while the [0, 0, 1]-ordered sBN breaks \{$C_{4z}$,  $C^{-1}_{4z}$, $M_x$, $M_y$, $T$, $TC_{2z}$, $TM_{xy}$, $TM_{-xy}$\} symmetries and belongs to the magnetic point group $4'm'm$.

\begin{figure}[h!]
\begin{center}
\fig{3.4in}{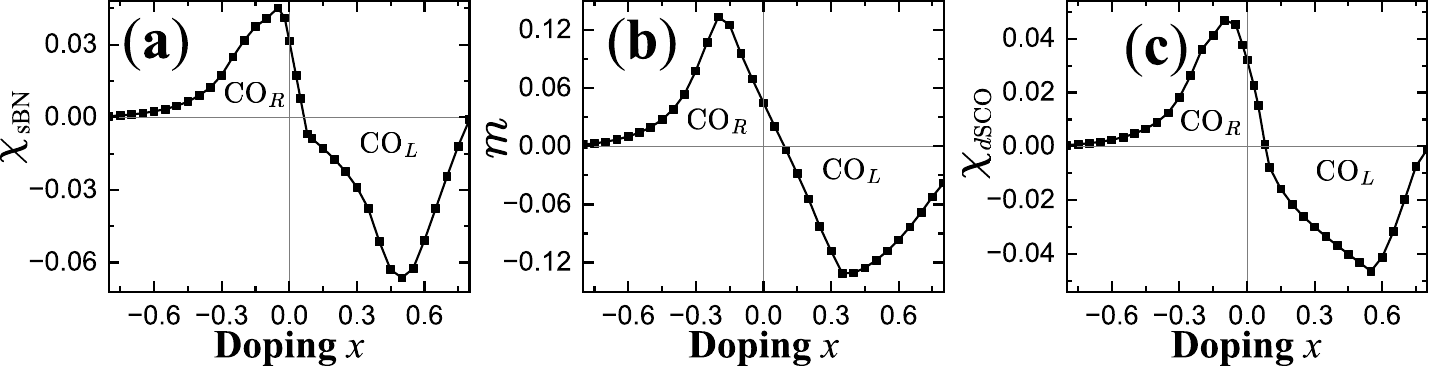}
\caption{The expectation of (a) sBN, (b) AFM, and (c) $d$SCO order as a function of doping concentration $x$ in the phenomenological coexistent state where the other two orders with the strength of 0.05 are introduced manually to the tight-binding model.}
\label{fig10}
\end{center}
\end{figure}

Using the symmetries of AFM, $d$SCO, and sBN listed in Table \ref{symm}, it is straightforward to obtain the symmetries of any coexistent state.
The coexistent state has a symmetry only if there exists a lattice translation that simultaneously recovers all involved states after the corresponding symmetry operation.
It is thus easy to show that all coexistent states, no matter it consists of any two or all three of AFM, $d$SCO, and sBN orders, are invariant under \{$E$, $C_{2z}$, $M_{xy}$, $M_{-xy}$\} operations, and belong to the magnetic group $mm2$.
Since they all possess the same symmetry properties, the combination of any two of these three orders would naturally leads to the emergence of the third one.
To demonstrate this, we have introduced phenomenologically two of these three orders, of the strength of 0.05, to the tight-binding Hamiltonian, and then calculate the expectation of the third order.
The doping dependence of the third order is shown in Fig. \ref{fig10}a for sBN, in Fig. \ref{fig10}b for AFM, and in Fig. \ref{fig10}c for $d$SCO.
Clearly, each of the three orders can be induced by the combination of the other two orders.
It varies continuously  as a function of doping concentration $x$, and changes its sign from negative to positive at a critical doping, indicating a transition in its chirality from left-handed to right-handed.

\bibliography{dPSCO2}

\end{document}